\documentclass[a4paper, amsfonts, amssymb, amsmath, reprint, showkeys, nofootinbib, twoside,notitlepage,onecolumn]{revtex4-2}

\def\apj{{ApJ}}                 

\def\prd{{Phys.~Rev.~D}}        
\def\prl{{Phys.~Rev.~Lett.}}    
\def\jqsrt{{J.~Quant.~Spec.~Radiat.~Transf.}}

\usepackage{amsmath,amstext}
\usepackage[T1]{fontenc}
\usepackage{amssymb}
\input{epsf}
\usepackage{graphicx}
\usepackage{ae,aecompl}

\usepackage{hyperref}
\usepackage{amsmath}
\usepackage{amssymb}
\usepackage{mathtools}
\usepackage{bm}
\usepackage{cleveref}
\usepackage{tensor}
\usepackage{braket}
\usepackage{enumitem}
\usepackage{mhchem}
\usepackage{cancel}
\usepackage{amsthm}
\usepackage{upgreek}
\usepackage{mathrsfs}
\usepackage{slashed}

\usepackage{color}

\newcommand{\spc}{\quad \quad \quad}

\newcommand{\q}{{\mathfrak{q}}}

\def\be{\begin{equation}}
\def\ee{\end{equation}}
\def\beq{\begin{eqnarray}}
\def\eeq{\end{eqnarray}}

\theoremstyle{definition}

\theoremstyle{theorem}
\newtheorem{theorem}{Theorem}

\theoremstyle{corollary}

\begin{document}
\title{Causality constraints on radiative transfer}
\author{ L.~Gavassino}
\affiliation{
Department of Mathematics, Vanderbilt University, Nashville, TN, USA
}

\begin{abstract}
The standard formula, due to Spiegel, for the smoothing of temperature fluctuations by radiative transfer is unstable in relativity. This is due to the fact that Spiegel neglected the transit time of light, thereby allowing the transport coefficients to move outside the convex geometry compatible with causality (the ``hydrohedron''). Here, we fix this pathology. First, we prove that the linearized radiative transfer equations are causal and covariantly stable by construction. Then, we repeat Spiegel's calculation accounting for the finite speed of photons. We find that the full transfer problem can be solved analytically. All the infinite (exact) transport coefficients arising from it fall inside the hydrohedron. Our analysis also accounts for isotropic scattering.
\end{abstract}

\maketitle

\section{Introduction}\label{Introne}

A recurrent question in relativistic hydrodynamics is how the principle of causality constrains the collective behavior of matter \cite{Bludman1968,Susskind1969,Bludman:1970at,Israel_Stewart_1979,Diener:1996mj,Adams:2006sv,Israel_2009_inbook,PuKoide2009fj,Hippert:2024hum,Hui:2025aja}. This question is of primary ``practical'' importance, since acausal hydrodynamic models become unstable under Lorentz boosts \cite{GavassinoSuperluminal2021}. Moreover, theorists are attracted to this topic because of the intriguing possibility of drawing a fundamental line between physically permitted and physically prohibited transport processes.

Unfortunately, the progress in this direction has been stagnant for decades. The reason is that hydrodynamics and causality seem to occupy opposite sides of the Fourier spectrum. In fact, hydrodynamics is an ``InfraRed'' (IR) property of the system, as it describes the large-scale behavior of matter, while causality is an ``UltraViolet'' (UV) feature, being mostly sensitive to the principal part of the equations (i.e. the highest derivatives). Making IR statements based on UV principles is usually quite hard. Indeed, even the question of whether the speed of sound should not exceed the speed of light has been debated for years \cite{Bludman1968,Susskind1969,Israel_2009_inbook,Camelio2022}. Fortunately, this fundamental IR-UV dichotomy has been recently resolved in two breakthrough papers \cite{HellerBounds2023,HellerHydroHedron2023jtd}. Combining both causality and stability requirements, the authors of \cite{HellerBounds2023,HellerHydroHedron2023jtd} were able to prove that, if one knows the value of the UV scale $\tau$ where hydrodynamics breaks down, then it is possible to rigorously constrain the IR, and derive an infinite list of inequalities on hydrodynamic transport coefficients in terms of $\tau$. Below, we summarize the main idea. 

Fix a macroscopic transport phenomenon, such as, for example, the diffusion of heat in a static medium. Then, in the spirit of an effective field theory, express the equation of motion (linearised around equilibrium) in the form of an infinite series in spatial gradients. In the case of heat diffusion along a single spatial direction, one may write
\begin{equation}\label{gradexp}
\partial_t  \delta T=D^{(1)}\partial_x^2 \delta T+D^{(3)}\partial_x^4 \delta T+D^{(5)}\partial_x^6 \delta T+... \, ,
\end{equation}
where $\delta T(t,x)$ is the local temperature fluctuation, and $D^{(2a-1)}$ are some constants, to be identified with the ``transport coefficients''. Finally, look for solutions to \eqref{gradexp} of the form $\delta T(t,x)=e^{ikx-i\omega t}$, with $k,\omega\in \mathbb{C}$. The result will be a dispersion relation $\omega(k)$, expressed as a Taylor series, whose radius of convergence $1/\tau$ marks the UV scale:
\begin{equation}\label{theseries}
\omega(k)=i\sum_{a=1}^{+\infty} (-1)^a D^{(2a-1)} k^{2a} \spc (\text{for }|k\tau|<1) \, .
\end{equation}
The key observation of \cite{HellerBounds2023,HellerHydroHedron2023jtd} is that, if equation \eqref{gradexp} is the formal gradient expansion of an underlying causal theory of matter near thermodynamic equilibrium, then it will necessarily obey the covariant stability condition\footnote{In \cite{HellerBounds2023}, the inequality $\mathfrak{Im}\,\omega\leq |\mathfrak{Im} \, k|$ is derived from the requirements that any retarded correlator be contained inside the future lightcone (causality) and be a tempered distribution (stability). In hydrodynamics, this same inequality is the statement that the fluid is stable in every reference frame \cite{GavassinoBounds2023,GavassinoDispersion2024}, meaning that the dispersion relation cannot be Lorenz-transformed into a growing Fourier mode \cite{Hiscock_Insatibility_first_order}.} $\mathfrak{Im}\,\omega\leq |\mathfrak{Im} \, k|$ for all $|k\tau|<1$. 
This can be translated into an infinite list of (rather complicated) universal inequalities on the dimensionless coefficients $D^{(2a-1)}/\tau^{2a-1}$. For illustration, here we report the constraints on the first two:
\begin{equation}\label{firsttwocondiHydroedron}
0 \leq \dfrac{D^{(1)}}{\tau} \leq \dfrac{16}{3\pi} \, , \spc -\dfrac{64}{15\pi} \leq \dfrac{D^{(3)}}{\tau^3} \leq \dfrac{256 +15\pi (8{-}3\pi D^{(1)}/\tau)D^{(1)}/\tau}{90\pi } \, .
\end{equation}
The authors of \cite{HellerHydroHedron2023jtd} also showed that these inequalities, combined together, force the infinite-dimensional vector $\{D^{(1)}/\tau,D^{(3)}/\tau^3,... \}$ to lie inside a convex set, which they called the ``hydrohedron''. Any hydrodynamic description of matter whose transport coefficients fall outside the hydrohedron is necessarily unstable in some Lorentz frame \cite{GavassinoBounds2023}, and unreliable in relativity.

Equipped with this new tool, we can now go back to known transport models, and simply ``check'' under which conditions they are consistent with relativity. For example, suppose that, in equation  \eqref{gradexp}, heat propagation is due exclusively to radiative transfer. Then, a calculation by \citet{Spiegel1957} (reported in textbooks \cite[\S 100]{mihalas_book}) tells us that\footnote{Note that, in the notation of \citet{Spiegel1957}, $\tau=\kappa^{-1}$. Furthermore, they express $\arctan(x)$ as $\text{arccot}(1/x)$.}
\begin{equation}\label{omspieg}
\begin{split}
\omega(k)={}& -i \dfrac{3D^{(1)}}{\tau^2} \bigg[1-\dfrac{\arctan(k\tau)}{k\tau} \bigg] \\ 
={}& i \dfrac{3D^{(1)}}{\tau^2} \sum_{a=1}^{+\infty} \dfrac{(-1)^a (k\tau)^{2a}}{1{+}2a}  \spc (\text{for }|k\tau|<1) \, , \\
\end{split}
\end{equation}
and we obtain, by comparison with \eqref{theseries}, the following infinite list of transport coefficients:
\begin{equation}\label{D2am1}
\dfrac{D^{(2a-1)}}{\tau^{2a-1}}=\dfrac{3}{1{+}2a} \dfrac{D^{(1)}}{\tau} \, ,
\end{equation}
which should fall into the hydrohedron for consistency with relativity.
Unfortunately, it can be shown that they never do so. The dispersion relation \eqref{omspieg} is \textit{always} unstable under some Lorentz boosts \cite{GavassinoDispRadiation2024cqw}, which implies that \eqref{omspieg} is incompatible causality for all non-zero values of $D^{(1)}/\tau$ \cite{GavassinoSuperluminal2021}. This is not so surprising, since Spiegel's derivation relies on a quasi-static approximation, which neglects the finite speed of propagation of photons. Thus, the question naturally arises as to how \eqref{omspieg} changes due to relativistic effects. Our goal, here, is to answer this question rigorously. 

Throughout the article, we work in Minkowski spacetime, with metric signature $(-,+,+,+)$, and adopt natural units $c=\hbar=k_B=1$. Greek indices ($\mu,\nu,...$) run from $0$ to $3$, while latine indices ($j,k,...$) run from 1 to 3.

\vspace{-0.4cm}
\section{Brief description of the instability of Spiegel's model}\label{iunstubuzzuz}
\vspace{-0.3cm}

Before discussing how to improve \eqref{omspieg}, let us discuss the origin and features of its instability in detail. Our notion of ``instability'' is the same as that of \citet{Hiscock_Insatibility_first_order}, namely the existence of growing Fourier modes in some reference frame, which signals the runaway of most solutions whose initial data belongs to $L^2(\mathbb{R}^3)$ in that frame.

\vspace{-0.4cm}
\subsection{Temperature runaways}\label{runaways}
\vspace{-0.2cm}

First of all, we note that \eqref{omspieg} is stable in the material rest frame. In fact,
the imaginary part of $\omega$ is negative for any $k\in \mathbb{R}/\{0\}$, which implies that every continuous superposition of ``Fourier modes'' (i.e. of modes with real $k$) is doomed to decay \cite{Hiscock_Insatibility_first_order}. But what about other reference frames? To answer this question, let $t'=\gamma(t{-}vx)$ and $x'=\gamma(x{-}tv)$ be the coordinates of an observer $\mathcal{O}'$, who moves with velocity $v$ relative to the medium (with $\gamma=1/\sqrt{1{-}v^2}$). Then, consider the plane wave $\delta T=e^{\Gamma't'/\tau}$, which describes a homogeneous temperature profile that changes only in time from the perspective of $\mathcal{O}'$. By relativity of simultaneity, the profile of $\delta T$ changes both in time \textit{and in space} in the rest frame of the medium, with associated wavenumbers
\vspace{-0.1cm}
\begin{equation}\label{hjhjhj}
\begin{cases}
\omega\tau =i\gamma\Gamma' \, ,\\
k\tau=i\gamma\Gamma' v\, .
\end{cases}
\end{equation}
Plugging these wavenumbers into \eqref{omspieg}, we obtain the following law:
\vspace{-0.1cm}
\begin{equation}
\gamma\Gamma'=  \dfrac{3D^{(1)}}{\tau} \bigg[\dfrac{\text{arctanh}(\gamma\Gamma'v)}{\gamma\Gamma'v} -1\bigg]\, .
\end{equation}
This is an implicit function, which expresses the growth rate $\Gamma'$ in terms of $v$. Solving it explicitly is out of the question. However, we can still write our solution as a parametric curve. Specifically, defined $\chi=\gamma \Gamma' v$, we have that
\vspace{-0.1cm}
\begin{equation}\label{brbrbrb}
\begin{cases}
\gamma\Gamma' = \dfrac{3D^{(1)}}{\tau} \bigg[\dfrac{\text{arctanh}(\chi)}{\chi} -1\bigg]\, , \\
\\
v=\dfrac{\chi}{\dfrac{3D^{(1)}}{\tau} \bigg[\dfrac{\text{arctanh}(\chi)}{\chi} -1\bigg]} \, .
\end{cases}
\end{equation}
Now we see the problem: If we send $\chi\rightarrow 1^-$, then $v\rightarrow 0$ and $\Gamma'\rightarrow +\infty$. This means that there exists a family of observers who move with \textit{infinitesimal} speeds relative to the medium, and who can witness spontaneous temperature runaways in the absence of any spatial gradients. Such sudden temperature explosions are reminiscent of the runaways of boosted parabolic theories \cite{Hiscock_Insatibility_first_order,Kost2000,GavassinoLyapunov_2020}, confirming that the instability is a consequence of causality violation \cite{GavassinoSuperluminal2021,GavassinoBounds2023}.

\subsection{Incompatibility with relativity}\label{AAA}

Let us now verify that the existence of runaway temperature profiles is associated with a breakdown of the inequality $\mathfrak{Im}\, \omega \leq |\mathfrak{Im}\, k|$, which makes Spiegel's model fundamentally incompatible with relativistic mechanics \cite{HellerBounds2023,HellerHydroHedron2023jtd,GavassinoBounds2023,GavassinoDispersion2024}. To this end, we combine \eqref{hjhjhj} with \eqref{brbrbrb}, and we obtain the following identity (valid for $\chi \in \mathbb{R}$):
\begin{equation}\label{donotworrione}
|\mathfrak{Im}\, (k\tau)|-\mathfrak{Im}\, (\omega\tau)= |\chi|+\dfrac{3D^{(1)}}{\tau} \bigg[1-\dfrac{\text{arctanh}(\chi)}{\chi} \bigg] \, .
\end{equation}
As $\chi$ approaches the unstable values (i.e. when $\chi\rightarrow 1^-$), the right-hand side of \eqref{donotworrione} diverges to $-\infty$. This in particular implies that the left side of \eqref{donotworrione} becomes negative, as claimed.




\subsection{Incompatibility with hydrohedron geometry}

We can finally show that the transport coefficients \eqref{D2am1} fall outside the hydrohedron \cite{HellerHydroHedron2023jtd}. To this end, we first note that the function \eqref{omspieg} coincides with its Taylor series within the circle $\mathfrak{C}=\{|k\tau|<1 \}$. It follows that the wavenumbers $k\tau= i(1^-)$, where the instability occurs (note that $k\tau=i\chi$), fall \textit{inside} the region where the gradient expansion \eqref{theseries} is applicable. Thus, some of the transport coefficients \eqref{D2am1} are incompatible with the analysis of \cite{HellerHydroHedron2023jtd}. Indeed, since the instability is caused by a negative divergence of \eqref{donotworrione}, we can argue that there must be an \textit{infinite number} of transport coefficients that exit the hydrohedron, for arbitrary choices of $D^{(1)}/\tau>0$.

\vspace{-0.2cm}
\section{The radiative transfer equations are causal and stable}
\vspace{-0.2cm}

Let us now discuss how to improve \eqref{omspieg}. The first step is to prove that the full radiative transfer theory is causal and covariantly stable. Of course, this should not come as a surprise, since the equation of radiative transfer is an ultra-relativistic Boltzmann equation \cite{Groot1980RelativisticKT,cercignani_book}, and the latter is known to be causal and stable. However, since the transport problem includes a coupling with the temperature of matter, it is important to have a complete proof.

\vspace{-0.2cm}
\subsection{Radiative transfer equations}
\vspace{-0.2cm}

Radiative transfer theory is concerned with the evolution of the temperature field $\delta T(x^\alpha)$ of a material medium, due to the ``random walk'' of photons traveling through it \cite{mihalas_book}. Usually, one assumes that the medium is in local thermodynamic equilibrium, and that it is sufficiently rigid/heavy to remain at rest at all times. The photons are modeled using kinetic theory. This means that one evolves the perturbation $ \delta f_\mathbf{p}(x^\alpha)$ to the invariant distribution function, which counts how many photons occupy a wavefunction located at the event $x^\alpha$ and having three-momentum $\mathbf{p}$. In the global rest frame of the medium, the linearized radiative transfer equations take the following form \cite{Pomraning1973,Minerbo1978,Levermore1984,CastorRadiationBook_2004}:
\begin{equation}\label{EoMsony}
\begin{split}
& c_M \partial_t \delta T +\partial_\mu \int \dfrac{2d^3 p}{(2\pi)^3 } p^\mu \delta f_\mathbf{p} =0 \, , \\
& \dfrac{p^\mu}{p^0}\partial_\mu \delta f_\mathbf{p} = \sigma_A \big(\delta f_{BB,p^0}-\delta f_\mathbf{p}\big)+\sigma_S \big(\delta \Bar{f}_{p^0} -\delta f_\mathbf{p}\big) \, . \\
\end{split}
\end{equation}
The first line is the conservation of energy for the total ``matter+radiation'' system, where $c_M$ is the specific heat of matter, and $p^\mu=(|\mathbf{p}|,\mathbf{p})$ is the photon four-momentum. The second line is the Boltzmann equation for the photon gas, where $\sigma_A(p^0)$ is the (energy-dependent) absorption coefficient, and $\sigma_S(p^0)$ is the (energy-dependent) scattering coefficient. The distribution function $f_{BB,p^0}=(e^{p^0/T}{-}1)^{-1}$ is the black-body occupation number, while the distribution function $\Bar{f}_{p^0}$ is the average of $f_\mathbf{p}$ over all directions at a given energy, so that
\begin{equation}\label{theotherfunctions}
\begin{split}
& \delta f_{BB,p^0} = f_{BB,p^0}(1{+}f_{BB,p^0}) \dfrac{p^0 \delta T}{T^2} \, , \\
& \delta \Bar{f}_{p^0}= \int \dfrac{d^2 \Omega}{4\pi} \delta f_{p^0\mathbf{\Omega}}\, . \\
\end{split}
\end{equation}
The absorption term in \eqref{EoMsony} owes its form to the Kirchhoff-Planck relation, which holds because the medium is assumed to be in local thermodynamic equilibrium. In the scattering term, we are assuming that each scattering process is isotropic. Note that the following identity holds by construction:
\begin{equation}\label{consturzionne}
\int \dfrac{2d^3p}{(2\pi)^3} \big(\delta \Bar{f}_{p^0} -\delta f_\mathbf{p}\big)Q(p^0)=0 \spc \big(\text{for any quantity }Q(p^0)\big) \, .
\end{equation}

\subsection{General proof of causality and covariant stability}

We are now ready to prove our first rigorous result. We will use the Gibbs stability criterion \cite{GavassinoGibbs2021,GavassinoStabilityCarter2022,GavassinoGENERIC2022} and the stability-causality correspondence \cite{GavassinoCausality2021,GavassinoSuperluminal2021} to prove that the radiative transfer theory is fully compatible with relativity. The precise statement is given below.
\begin{theorem}\label{Theo1}
If $T>0$, $c_M>0$, $\sigma_A(p^0)\geq 0$, and $\sigma_S(p^0)\geq 0$, the system \eqref{EoMsony} is causal and covariantly stable.
\end{theorem}
\begin{proof}
The information current $E^\mu$ of the total ``matter+radiaton'' system is just the sum of the matter part and the radiation part, namely \cite{GavassinoGibbs2021,GavassinoCausality2021,GavassinoNonHydro2022,RochaGavassinoFluctuKin2024afv}
\begin{equation}
E^\mu =c_M \dfrac{(\delta T)^2}{T^2} \dfrac{u^\mu}{2} +\dfrac{1}{2} \int \dfrac{2d^3 p}{(2\pi)^3 p^0}  \dfrac{(\delta f_\mathbf{p})^2 p^\mu}{f_{BB,p^0}(1{+}f_{BB,p^0})} \, ,
\end{equation}
with $u^\mu=(1,0,0,0)$.
Clearly, $E^\mu$ is timelike future-directed and non-vanishing for any choice of couple $\{\delta T,\delta f_\mathbf{p} \}\neq 0$. If we take its divergence, and invoke \eqref{EoMsony}, we recover, with the aid of \eqref{consturzionne}, the second law of thermodynamics:
\begin{equation}\label{secondlaw}
\partial_\mu E^\mu = -\int \dfrac{2d^3p}{(2\pi)^3} \bigg[ \dfrac{\sigma_A\big(\delta f_{BB,p^0}-\delta f_\mathbf{p}\big)^2}{f_{BB,p^0}(1{+}f_{BB,p^0})} +\dfrac{\sigma_S \big(\delta \Bar{f}_{p^0} -\delta f_\mathbf{p}\big)^2}{f_{BB,p^0}(1{+}f_{BB,p^0})}\bigg] \leq 0 \, .
\end{equation}
Then, the results in \cite{GavassinoGibbs2021,GavassinoCasmir2022} apply, and the evolution equations are automatically causal and covariantly stable.
\end{proof}

\subsection{Plane-wave causality and stability analysis}

Given the statement of Theorem \ref{Theo1}, it should be clear that any dispersion relation arising from \eqref{EoMsony} must obey the inequality $\mathfrak{Im}\,\omega\leq |\mathfrak{Im} \, k|$, since the latter is ultimately equivalent to covariant stability itself \cite{GavassinoBounds2023}. Nevertheless, it is instructive to set up a more direct proof, which we provide below.

In the system \eqref{EoMsony}, plug the second equation into the first, and assume that $\{\delta T,\delta f_\mathbf{p} \}\propto e^{ikx-i\omega t}$. The result is
\begin{equation}\label{EoMFour}
\begin{split}
-i c_M \omega \delta T={}& -\int \dfrac{2d^3 p}{(2\pi)^3 } p^0 \big[ \sigma_A \big(\delta f_{BB,p^0}-\delta f_\mathbf{p}\big)+\sigma_S \big(\delta \Bar{f}_{p^0} -\delta f_\mathbf{p}\big)\big]  \, , \\
-i\bigg(\omega {-}k\dfrac{p^1}{p^0} \bigg) \delta f_\mathbf{p} ={}& \sigma_A \big(\delta f_{BB,p^0}-\delta f_\mathbf{p}\big)+\sigma_S \big(\delta \Bar{f}_{p^0} -\delta f_\mathbf{p}\big) \, . \\
\end{split}
\end{equation}
Then, we have the following statement:
\begin{theorem}\label{theo2}
If $T>0$, $c_M>0$, $\sigma_A(p^0)\geq 0$, and $\sigma_S(p^0)\geq 0$, all solutions of \eqref{EoMFour} fulfill the inequality $\mathfrak{Im}\, \omega \leq |\mathfrak{Im}\, k|$.
\end{theorem}
\begin{proof}
The main idea is to ``reconstruct'' a plane-wave version of equation \eqref{secondlaw}. To this end, we multiply the first equation of \eqref{EoMFour} by $\delta T^*/T^2$ and the second by $\delta f_\mathbf{p}^*/[f_{BB,p^0}(1{+}f_{BB,p^0})]$, and integrate the second over all $\mathbf{p}$, giving
\begin{equation}\label{Emu2}
\begin{split}
-i c_M \omega \dfrac{|\delta T|^2}{T^2} ={}& -\int \dfrac{2d^3 p}{(2\pi)^3 } \delta f_{BB,p^0}^* \dfrac{\sigma_A \big(\delta f_{BB,p^0}-\delta f_\mathbf{p}\big)+\sigma_S \big(\delta \Bar{f}_{p^0} -\delta f_\mathbf{p}\big)}{f_{BB,p^0}(1{+}f_{BB,p^0})} \, , \\
-i\int \dfrac{2d^3 p}{(2\pi)^3 }\bigg(\omega {-}k\dfrac{p^1}{p^0} \bigg) \dfrac{|\delta f_\mathbf{p}|^2}{f_{BB,p^0}(1{+}f_{BB,p^0})} ={}& \int \dfrac{2d^3 p}{(2\pi)^3 } \delta f_\mathbf{p}^* \,  \dfrac{ \sigma_A \big(\delta f_{BB,p^0}-\delta f_\mathbf{p}\big)+\sigma_S \big(\delta \Bar{f}_{p^0} -\delta f_\mathbf{p}\big) }{f_{BB,p^0}(1{+}f_{BB,p^0})} \, . \\
\end{split}
\end{equation}
Adding these two equations, we obtain the desired analog of \eqref{secondlaw}. We can then isolate $\omega$, and we obtain
\begin{equation}
\omega =k \dfrac{\displaystyle\int \dfrac{2d^3 p}{(2\pi)^3 }  \dfrac{|\delta f_\mathbf{p}|^2 } {f_{BB,p^0}(1{+}f_{BB,p^0})} \dfrac{p^1}{p^0}}{c_M \dfrac{|\delta T|^2}{T^2} + \displaystyle\int \dfrac{2d^3 p}{(2\pi)^3 }  \dfrac{|\delta f_\mathbf{p}|^2 }{f_{BB,p^0}(1{+}f_{BB,p^0})}} -i \dfrac{\displaystyle\int \dfrac{2d^3p}{(2\pi)^3} \bigg[ \dfrac{\sigma_A|\delta f_{BB,p^0}-\delta f_\mathbf{p}|^2}{f_{BB,p^0}(1{+}f_{BB,p^0})} +\dfrac{\sigma_S |\delta \Bar{f}_{p^0} -\delta f_\mathbf{p}|^2}{f_{BB,p^0}(1{+}f_{BB,p^0})}\bigg]}{c_M \dfrac{|\delta T|^2}{T^2} + \displaystyle\int \dfrac{2d^3 p}{(2\pi)^3 }  \dfrac{|\delta f_\mathbf{p}|^2 }{f_{BB,p^0}(1{+}f_{BB,p^0})}} \, .
\end{equation}
Taking the imaginary part of this equation, it is immediate to see that, indeed, $\mathfrak{Im}\,\omega\leq |\mathfrak{Im} \, k|$.
\end{proof}
The above theorem implies that, if one were able to compute the \textit{exact} dispersion relations $\omega(k)$ of radiative transfer, avoiding the quasistatic approximation adopted by \citet{Spiegel1957}, the resulting model would automatically be causal and covariantly stable, and the associated transport coefficients $D^{(2a-1)}$ would automatically fall within the hydrohedron. 

\section{Exact dispersion relations}

Now that we have proven that the radiative transfer theory (when solved \textit{exactly}) is fully consistent with relativity, we can compute the relativistic analog of equation \eqref{omspieg}.

\subsection{Complete calculation in full generality}

Our starting point is the second line in equation \eqref{EoMFour}. Isolating $\delta f_\mathbf{p}$, we obtain the following identity:
\begin{equation}\label{guyone}
\delta f_\mathbf{p} =\dfrac{\sigma_A \delta f_{BB,p^0}+\sigma_S \delta \Bar{f}_{p^0}}{\sigma_A +\sigma_S -i\omega +ikp^1/p^0} \, .
\end{equation}
The risk of dividing by zero is not a big problem, because $p^1/p^0$ is a continuous parameter, and $\delta f_\mathbf{p}$ is a distribution over $\mathbf{p}$. This means that the value of $\delta f_\mathbf{p}$ is allowed to be singular for some $p^1/p^0$, as long as the integrals in $\mathbf{p}$ remain finite. 
Plugging \eqref{guyone} into the second line of \eqref{theotherfunctions}, we get an equation for $\delta \Bar{f}_{p^0}$ in terms of $\delta f_{BB,p^0}$, whose solution is
\begin{equation}
\begin{split}
\delta \Bar{f}_{p^0}={}& \dfrac{H \sigma_A \delta f_{BB,p^0}}{1-H\sigma_S} \, , \\
H \equiv {}& \int \dfrac{d^2 \Omega}{4\pi} \dfrac{1}{\sigma_A {+}\sigma_S {-}i\omega {+}ik\Omega^1}= \dfrac{1}{k} \arctan\bigg(\dfrac{k}{\sigma_A {+}\sigma_S {-}i\omega} \bigg) \, . \\
\end{split}
\end{equation}
Note that $H$ in general depends on $p^0$, since $\sigma_A$ and $\sigma_S$ are functions of $p^0$.
We can insert the above expression for $\delta \Bar{f}_{p^0}$ back into \eqref{guyone}, and use the first equation of \eqref{theotherfunctions} to express $\delta f_\mathbf{p}$ solely in terms of $\delta T$:
\begin{equation}
\delta f_\mathbf{p} =\dfrac{\sigma_A f_{BB,p^0}(1{+}f_{BB,p^0}) p^0/T^2}{(\sigma_A +\sigma_S -i\omega +ikp^1/p^0)(1-H\sigma_S)} \, \delta T \, .
\end{equation}
Plugging this equation into the first line of \eqref{EoMsony}, the factor $\delta T$ cancels out, and we obtain an \textit{exact} implicit formula for $\omega$ as a function of $k$:
\begin{equation}\label{milosch}
\omega=-i \dfrac{c_R}{c_M} \bigg\langle \sigma_A \dfrac{1-\dfrac{\sigma_A{+}\sigma_S}{k}\arctan\bigg(\dfrac{k}{\sigma_A{+}\sigma_S{-}i\omega}\bigg)}{1-\dfrac{\sigma_S}{k}\arctan\bigg(\dfrac{k}{\sigma_A{+}\sigma_S{-}i\omega}\bigg)} \bigg\rangle \, ,
\end{equation}
where $c_R=4a_RT^3$ is the specific heat of radiation, with $a_R=\pi^2/15$ the usual radiation constant. The operation $\langle ... \rangle$ denotes the average over all energies, with ordinary Rosseland weight $(p^0)^4 f_{BB,p^0}(1{+}f_{BB,p^0})$ \cite[\S 6.7]{CastorRadiationBook_2004}.

\subsection{Grey material}

The original calculation by \citet{Spiegel1957} was carried out under the assumption of a grey material, i.e. a substance such that $\sigma_A$ and $\sigma_S$ do not depend on $p^0$. In this case, one can drop the average symbol in \eqref{milosch}. Then, defining the radiation dominance $\lambda$, the photon mean free path $\tau$, and the single-scattering albedo $\varpi$ \cite[\S 5.2]{CastorRadiationBook_2004}, respectively,
\begin{equation}
\lambda =\dfrac{c_R}{c_M} \, , \spc \tau =\dfrac{1}{\sigma_A {+}\sigma_S} \, , \spc \varpi =\dfrac{\sigma_S}{\sigma_A {+}\sigma_S} \, ,
\end{equation}
equation \eqref{milosch} acquires the following simpler form:
\begin{equation}\label{roku}
\omega=-i \dfrac{\lambda}{\tau}  (1{-}\varpi) \, \dfrac{1-\dfrac{1}{k\tau}\arctan\bigg(\dfrac{k\tau}{1{-}i\omega \tau}\bigg)}{1-\dfrac{\varpi}{k\tau}\arctan\bigg(\dfrac{k\tau}{1{-}i\omega\tau}\bigg)}  \, .
\end{equation}
This implicit formula for $\omega(k)$ is the relativistic generalization of \eqref{omspieg}, which accounts also for isotropic scattering (through $\varpi$). It is exact, and its solutions are guaranteed to be covariantly stable, by Theorem \ref{theo2}. As a result, the radiative transport coefficients that arise from it always fall into the hydrohedron, for any $\lambda>0$, $\tau>0$, and $\varpi \in [0,1]$.

In Appendix \ref{parametrizzone}, we show that all solutions to \eqref{roku} can be expressed explicitly as parametric curves.

\section{Consequences of relativity}

We are finally entering the central part of the paper, where we discuss how the principle of causality modifies equation \eqref{omspieg}. In the following calculations, we will assume for simplicity that scattering is absent (i.e. $\varpi=0$). It should be clear at this point that our main conclusions do not change qualitatively when $\varpi \neq 0$.

\subsection{Causal delays prevent instabilities}

When $\varpi=0$, the fully relativistic dispersion relation \eqref{roku} is very similar to \eqref{omspieg}, except for a denominator ``$1{-}i\omega\tau$'' in the argument of the inverse tangent:
\begin{equation}\label{fullrelat}
 \omega=-i \dfrac{\lambda}{\tau} \bigg[ 1-\dfrac{1}{k\tau}\arctan\bigg(\dfrac{k\tau}{1{-}i\omega \tau}\bigg) \bigg] \, .
\end{equation}
This correction is responsible for causality restoration. In fact, it can be verified (see Appendix \ref{time retardation}) that the existence of such a term is related to the presence of a retarded time in the ``energy propagator'', which ensures that the heat transported by photons travels on the surface of the lightcone, and not outside of it. This automatically heals the model from the instability described in section \ref{iunstubuzzuz} (as stated in Theorem \ref{theo2}).

\begin{figure}[h!]
    \centering
    \includegraphics[width=0.55\linewidth]{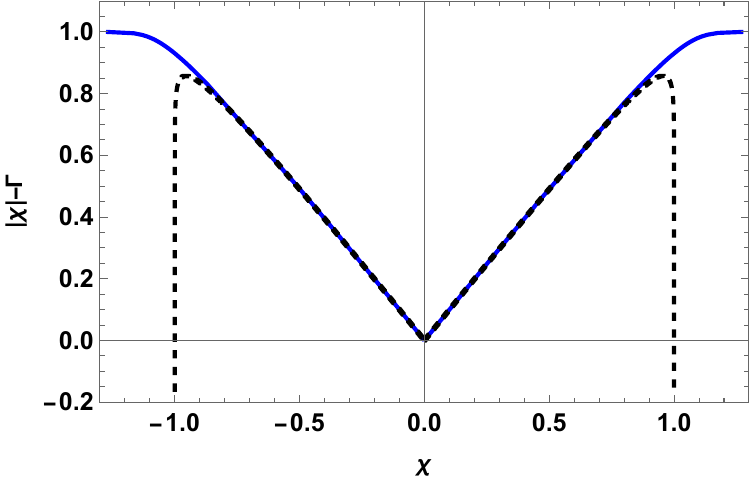}
\caption{Graph of the covariant-stability discriminant $|\chi|-\Gamma$ as a function of $\chi$, for $\lambda=0.1$. If the discriminant becomes negative, radiative transfer is unstable to some observers. The dashed line represents Spiegel's model \eqref{omspieg}, which becomes unstable near $\chi=\pm 1$ (for any $\lambda$). The blue line is the discriminant of the exact dispersion relation \eqref{fullrelat}, which is covariantly stable by Theorem \ref{theo2}. As can be seen, no pathology arises in \eqref{fullrelat} around $\chi=\pm 1$. Furthermore, it can be proven that, as $\chi\rightarrow \infty$, the discriminant of \eqref{fullrelat} asymptotically approaches 1. The same qualitative behavior is observed for all positive $\lambda$.}
    \label{fig:stiliamo}
\end{figure}

We can verify this more directly. To this end, let us first note that the ``dangerous'' wavenumbers considered in equation \eqref{hjhjhj} are purely imaginary. Hence, defined $\chi=-ik\tau$, $\Gamma=-i\omega\tau$, and $r=\chi/(1{+}\Gamma)$, let us impose that $\chi,\Gamma,r \in \mathbb{R}$. Then, equation \eqref{fullrelat} defines a parametric curve $\{\chi(r),\Gamma(r) \}:\mathbb{R}\rightarrow \mathbb{R}^2$, which reads as follows:
\begin{equation}
\begin{cases}
\Gamma= -\dfrac{1{+}\lambda}{2}+\sqrt{\bigg(\dfrac{1{-}\lambda}{2} \bigg)^2+\lambda \dfrac{\text{arctanh}(r)}{r}} \, , \\
\chi=r\bigg[ \dfrac{1{-}\lambda}{2}+\sqrt{\bigg(\dfrac{1{-}\lambda}{2} \bigg)^2+\lambda \dfrac{\text{arctanh}(r)}{r}}\bigg] \, . \\
\end{cases}
\end{equation}
In Figure \ref{fig:stiliamo}, we plot the covariant-stability discriminant $|\mathfrak{Im}\, (k\tau)|-\mathfrak{Im}\, (\omega\tau)=|\chi|-\Gamma$, which, we recall, must be non-negative in relativistic systems near equilibrium \cite{HellerBounds2023,GavassinoBounds2023}, but fails to be such in Spiegel's model (see section \ref{AAA}). We clearly see that the presence of the time retardation term $1{-}i\omega\tau$ has healed all pathologies, making \eqref{fullrelat} a viable model for energy transport in relativity.


\subsection{Causality reduces the transport coefficients}

The time-delay correction $1-i\omega\tau$ in \eqref{fullrelat} sets a causal bound on the transport of energy, and this reduces the magnitude of all transport coefficients $D^{(2a-1)}$, forcing them to fall into the hydrohedron (again, by Theorem \ref{theo2}). Unfortunately, there is no single formula that provides all exact coefficients $D^{(2a-1)}$ in full relativity, because the expression \eqref{fullrelat} is an implicit function. However, using the implicit function theorem, one can recursively evaluate each $D^{(2a-1)}$ one by one. In the table below, we provide the first four, together with their small and large $\lambda$ limits.

\begin{table}[h!]
    \centering
    \begin{tabular}{|c|c|c|c|}
\hline
Coefficient & $\quad$ Spiegel \cite{Spiegel1957} ($\equiv$ Small $\lambda$) $\quad$ & Full Relativity & $\quad$ Large $\lambda$  $\quad$  \\
\hline
\hline
& & & \\[-1em]
$\dfrac{D^{(1)}}{\tau}$   & $\dfrac{\lambda}{3}$ & $\dfrac{\lambda}{3(1+\lambda)}$ & $\dfrac{1}{3}$ \\
& & & \\[-1em]
\hline
& & & \\[-1em]
$\dfrac{D^{(3)}}{\tau^3}$   & $\dfrac{\lambda}{5}$ & $\dfrac{\lambda  \left(9+3\lambda-\lambda ^2\right)}{45 (1+\lambda)^3}$ & $-\dfrac{1}{45}$ \\
& & & \\[-1em]
\hline
& & & \\[-1em]
$\dfrac{D^{(5)}}{\tau^5}$   & $\dfrac{\lambda}{7}$ & $\dfrac{\lambda  \left(135+36 \lambda-51 \lambda ^2-20 \lambda ^3+2 \lambda ^4 \right)}{\textcolor{white}{\dfrac{1}{1}} 945 (1+\lambda)^5\textcolor{white}{\dfrac{1}{1}}}$ & $\dfrac{2}{945}$ \\
\hline
& & & \\[-1em]
$\dfrac{D^{(7)}}{\tau^7}$   & $\dfrac{\lambda}{9}$ & $\dfrac{\lambda  \left(1575-135 \lambda-1593 \lambda ^2-787 \lambda ^3+37 \lambda ^4+63 \lambda ^5 -3 \lambda ^6 \right)}{14175 (1+\lambda)^7\textcolor{white}{\dfrac{1}{1}}}$  & $-\dfrac{1}{4725}$ \\
\hline
\end{tabular}
\end{table}

It can be verified from the table above that each fully relativistic $D^{(2a-1)}$ is smaller than its Spiegel's correspondent. Indeed, some transport coefficients even become negative. Specifically, we find that, as $a$ grows, each function $D^{(2a-1)}(\lambda)$ undergoes more and more ``sign fluctuations''. For example, while $D^{(1)}$ is always positive, $D^{(3)}$ becomes negative at $\lambda=3(1{+}\sqrt{5})/2$, and $D^{(5)}$ becomes negative at $\lambda\approx 1.57$ and again positive at $\lambda \approx 12$. The coefficient $D^{(17)}$ changes sign several times, at $\lambda \approx  0.199$, $0.614$, $1.45$, $3.11$, $6.35$, $12.9$, $28.9$, and $95.8$. 

A notable implication of the change of sign of $D^{(3)}$ is that the Super-Burnett truncation \cite{Shavaliyev1993}, namely
\begin{equation}
\partial_t  \delta T=\dfrac{\lambda \tau}{3(1{+}\lambda)}\bigg[\partial_x^2 \delta T+\dfrac{  9{+}3\lambda{-}\lambda ^2}{15 (1{+}\lambda)^2} \tau^2 \partial_x^4 \delta T\bigg] \, ,
\end{equation}
is unstable in the rest frame for $\lambda<3(1{+}\sqrt{5})/2$, but becomes rest-frame stable at larger radiation dominance.

\begin{figure}[h!]
    \centering
    \includegraphics[width=0.55\linewidth]{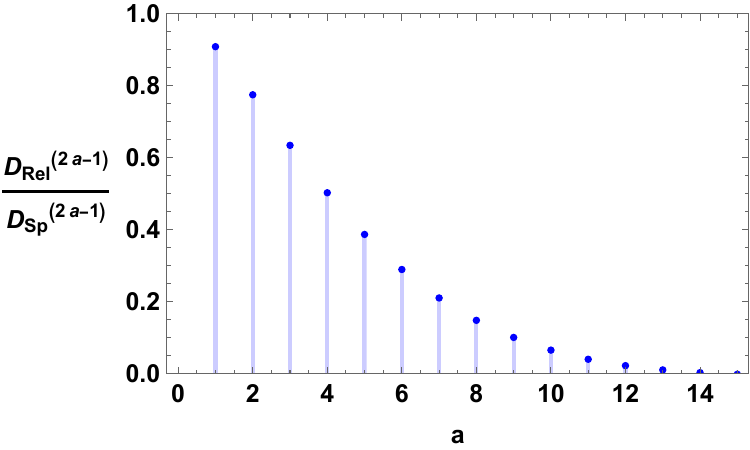}
\caption{Ratios $D^{(2a-1)}_{\text{Rel}}/D^{(2a-1)}_{\text{Sp}}$, where $D^{(2a-1)}_{\text{Rel}}$ are the relativistic transport coefficients computed from \eqref{fullrelat}, while $D^{(2a-1)}_{\text{Sp}}$ are computed from Spiegel's approximation \eqref{omspieg}. The rate of decay of the graph decreases with $\lambda$ (here, we chose $\lambda=0.1$).}
    \label{fig:DasaGrows}
\end{figure}

Following the same procedure as in \cite[Sec. II.E]{GavassinoDispRadiation2024cqw}, one can easily prove that Spiegel's model \eqref{omspieg} is the truncation of \eqref{fullrelat} to first order in $\lambda$. It follows that, if we fix $a$, and take the small-$\lambda$ limit, the relativistic transport coefficient $D^{(2a-1)}(\lambda)$ reduces to Spiegel's transport coefficient \eqref{D2am1}. However, if we fix $\lambda$, and take the large-$a$ limit, Spiegel's transport coefficients become infinitely larger than the relativistic ones (see Figure \ref{fig:DasaGrows}). This different scaling of $D^{(2a-1)}$ with $a$ is ultimately what causes infinite coefficients to fall outside the hydrohedron in Spiegel's model.

\subsection{Radiation-dominated systems}

\begin{figure}[h!]
    \centering
    \includegraphics[width=0.47\linewidth]{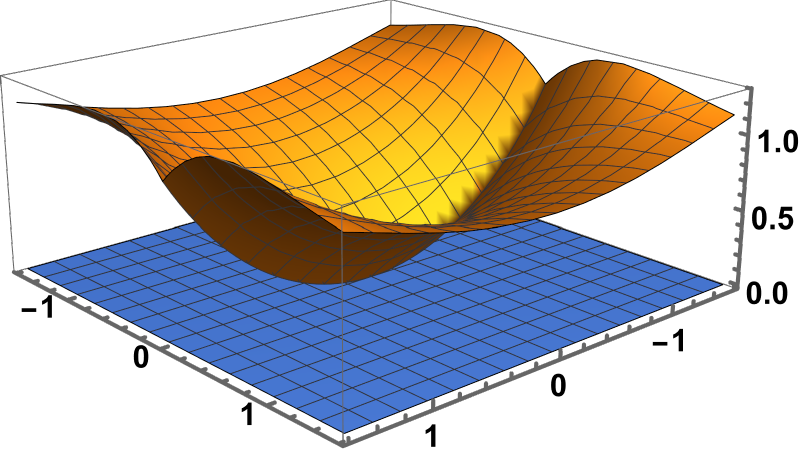}
    \caption{Graph of the covariant-stability discriminant $|\mathfrak{Im} (k\tau)|-\mathfrak{Im}(\omega \tau)$ as a function $k\tau \in \mathbb{C}$ for a radiation-dominated system. The blue plane marks level 0. The discriminant never cuts below the plane, showing that the dispersion relation \eqref{panona} is covariantly stable.}
    \label{fig:radomone}
\end{figure}

As can be seen from the table in the previous section, Spiegel's quasi-static approximation of the transport coefficients is especially off in high-temperature systems, where the specific heat is dominated by the radiative component (i.e. $\lambda \rightarrow \infty$). The reason is that, in this limit, the temperature evolves over a timescale $\omega^{-1}$ that is comparable to the photon time of flight $\tau$, making time-retardation effects very large. The dispersion relation \eqref{fullrelat} then becomes
\begin{equation}\label{panona}
\omega(k)\approx - \dfrac{i}{\tau} \bigg[ 1-\dfrac{k\tau}{\tan(k\tau)} \bigg] \spc (\text{for }|k\tau|<\pi/2) \, .
\end{equation}
To derive \eqref{panona}, one can just assume that the left side of \eqref{fullrelat} be negligible when $\lambda$ is large, and thus set the quantity in square brackets to 0 (a more formal derivation involves sending $\lambda\to \infty$ in \eqref{koM}).
Despite being an approximation, 
the high-temperature formula \eqref{panona} fully accounts for all causal bounds on the motion of light, and it is therefore covariantly stable, see figure \ref{fig:radomone}. Indeed, we note that equation \eqref{panona} coincides with the dispersion relation of relativistic kinetic theory in the Relaxation Time Approximation (RTA) \cite{Romatschke:2015gic,Brants:2024wrx}. This is no surprise, since the equations of radiative transfer \eqref{EoMsony} reduce to an RTA model for the photon gas, in the limit where $c_M \rightarrow 0$ and $\sigma_S\rightarrow 0$.

\section{Conclusions}

As expected, the radiative transfer equations give rise to a perfectly causal and stable relativistic theory of heat conduction. This implies that the instability of Spiegel's model is a mere artifact of the quasistatic approximation. Indeed, the decision to neglect the travel time of photons is unacceptable in relativity, due to the deep mathematical connection between causality and covariant stability \cite{GavassinoSuperluminal2021,GavassinoBounds2023}, which makes acausal models (like Spiegel's one) unusable when boosted. Not surprisingly, if one properly accounts for relativistic time delays, the resulting \textit{exact} dispersion relation \eqref{roku} (which also accounts for scattering) is perfectly well-behaved in every reference frame. 

In the absence of scattering, the causal retardation converts equation \eqref{omspieg} into \eqref{fullrelat}. On the surface, the causal delay (i.e. the denominator $1-i\omega\tau$ in the inverse tangent) seems to be a minor correction. Indeed, if $|\omega\tau|\ll 1$, Spiegel's model is quite accurate. However, when $\omega$ becomes of the order of $1/\tau$, relativistic effects arise, which act to ``contain'' the growth of $\omega$, preventing the formation of unstable waves at large $k\tau$ (see figure \ref{fig:stiliamo}). 

A notable situation where $\omega$ quickly becomes comparable to $\tau$ is when the specific heat of radiation is much larger than that of matter (i.e., in radiation-dominated systems). In this limit, Spiegel's model is never applicable, even at relatively small $k\tau$, and the system transitions to a new behavior, where the dispersion relation is now given by \eqref{panona}. The latter formula coincides with the diffusive mode of an ideal massless gas in the relaxation time approximation, whose transport coefficients admit an explicit representation in terms of Riemann's Zeta function:
\begin{equation}
\dfrac{D^{(2a-1)}}{\tau^{2a-1}} =(-1)^{a+1} \dfrac{2\, \zeta(2a)}{\pi^{2a}} \, .
\end{equation}
Comparison with \eqref{D2am1} shows that the causal delay has modified the dependence of $D^{(2a-1)}$ on $a$ in a dramatic way. Specifically, the sequence $D^{(2a-1)}$ now decays like $1/\pi^{2a}$ at large $a$, while it decayed like $1/a$ in \eqref{D2am1}. Moreover, the causal delay has changed the \textit{sign} of all transport coefficients with $a$ even. 


\newpage
\section*{Acknowledgements}

This work is partially supported by a Vanderbilt's Seeding Success Grant.

\appendix

\section{Parametric representation of the dispersion relation}\label{parametrizzone}

\begin{figure}[h!]
    \centering
    \includegraphics[width=0.55\linewidth]{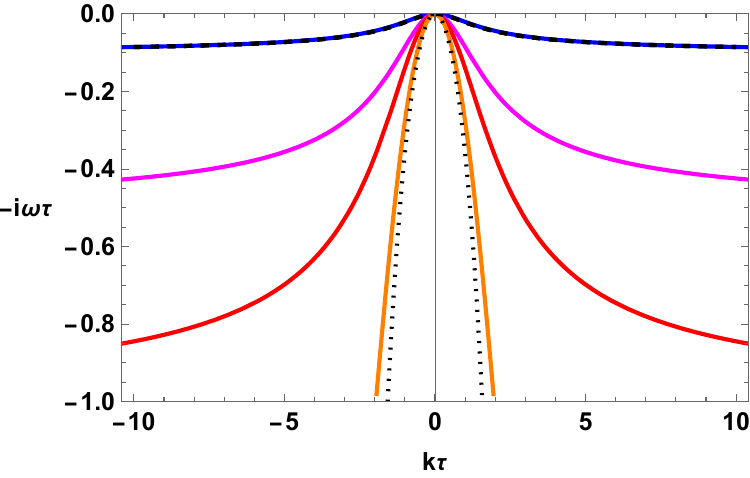}
    \caption{Graph of the scatteringless dispersion relation \eqref{fullrelat} [or equivalently \eqref{koM}] for different values of the radiation dominancy $\lambda$, respectively $0.1$ (blue), $0.5$ (magenta), $1$ (red), and $5$ (orange). The dashed line is Spiegel's model for $\lambda=0.1$, and the dotted line is the limiting formula \eqref{panona}.}
    \label{fig:Allregimes}
\end{figure}

We start from equation \eqref{roku}, and we define the dimensionless variables
$\Gamma=-i\omega\tau$ and $\q=k\tau$. In these units, the dependence on $\tau$ effectively disappears, and we have
\begin{equation}\label{kviate}
\Gamma= -\lambda  (1{-}\varpi) \, \dfrac{1-\dfrac{1}{\q}\arctan\bigg(\dfrac{\q}{1{+}\Gamma}\bigg)}{1-\dfrac{\varpi}{\q}\arctan\bigg(\dfrac{\q}{1{+}\Gamma}\bigg)}  \, .
\end{equation}
Now, let us introduce the parameter $r=\q/(1+\Gamma)$, and let us use it to replace $\q$ everywhere in \eqref{kviate}. The result is a quadratic equation for $\Gamma$ in terms of $r$, which can be solved analytically. This also allows us to express $\q$ in terms of $r$, since $\q=r(1+\Gamma)$. At the end of the day, we have the following exact parametric curve:
\begin{equation}
\begin{split}
k(r)={}&\dfrac{r}{\tau} \bigg[\frac{1}{2}  (\lambda  \varpi -\lambda +1)+\frac{\sqrt{\left(\lambda  r (\varpi -1)+\varpi  \arctan(r)+r\right)^2-4 r (\lambda  (\varpi -1)+\varpi ) \arctan (r)}}{2r}+\frac{\varpi  \arctan(r)}{2r} \bigg] \, , \\
\omega(r)={}&\dfrac{i}{\tau} \bigg[ \frac{1}{2} (\lambda  \varpi -\lambda -1)+\frac{\sqrt{\left(\lambda  r (\varpi -1)+\varpi  \arctan (r)+r\right)^2-4 r (\lambda  (\varpi -1)+\varpi ) \arctan(r)}}{2 r}+\frac{\varpi  \arctan (r)}{2 r} \bigg] \, . \\
\end{split}
\end{equation}
In the case where $\varpi=0$, the above formula simplifies considerably:
\begin{equation}\label{koM}
\begin{split}
k(r)={}& \dfrac{r}{\tau} \bigg[ \dfrac{1{-}\lambda}{2} +\sqrt{\bigg(\dfrac{1{-}\lambda}{2} \bigg)^2 +\lambda \dfrac{\arctan(r)}{r}} \bigg] \, , \\
\omega(r)={}& -\dfrac{i}{\tau}  \bigg[ \dfrac{1{+}\lambda}{2} -\sqrt{\bigg(\dfrac{1{-}\lambda}{2} \bigg)^2 +\lambda \dfrac{\arctan(r)}{r}} \bigg] \, . \\
\end{split}
\end{equation}
This parametric expression allows us to easily plot the dispersion relation \eqref{fullrelat} for different values of $\lambda$ (see figure \ref{fig:Allregimes}).

\section{Time retardation}\label{time retardation}

In this appendix, we show that (for grey ``scatteringless'' media) the only approximation in the original model by \citet{Spiegel1957} was neglecting a time retardation in the expression for the radiative stress-energy tensor. This time retardation, when properly included, gives rise to the denominator $1-i\omega\tau$ in equation \eqref{fullrelat}. 

\vspace{-0.2cm}
\subsubsection{Exact solution of the radiative Boltzmann equation}
\vspace{-0.2cm}

In the absence of scattering (i.e. $\sigma_S=0$) and for grey media (i.e. $\sigma_A=\tau^{-1}=\text{const}$), the second equation of \eqref{EoMsony} becomes
\begin{equation}\label{boltzmann}
\dfrac{p^\mu}{p^0}\partial_\mu  \delta f_\mathbf{p} +\tau^{-1} \delta f_\mathbf{p} =  f_{BB,p^0}(1{+}f_{BB,p^0}) \dfrac{p^0 \delta T}{T^2\tau} \, .
\end{equation}
This expression can be converted into an ordinary differential equation. In fact, if we travel along the photon's worldline $x^\mu(\lambda)=x^\mu-\lambda p^\mu/p^0$, and we regard $\delta f$ and $\delta T$ as functions of $\lambda$, the derivative $\frac{p^\mu}{p^0}\partial_\mu$ becomes $-d/d\lambda$. Hence, equation \eqref{boltzmann} can be analytically solved by multiplying both sides by the integrating factor $e^{-\lambda/\tau}$, giving
\begin{equation}\label{granch}
\delta f(x^\mu)= f_{BB,p^0}(1{+}f_{BB,p^0}) \dfrac{p^0 }{T^2}\int_0^{+\infty} \delta T\bigg(x^\mu {-}\lambda \dfrac{p^\mu}{p^0} \bigg) \dfrac{e^{-\lambda/\tau}}{\tau} d\lambda \, .   
\end{equation}
Note that, if $\tau \rightarrow 0$, equation \eqref{granch} reduces to $\delta f_\mathbf{p}=\delta f_{BB,p^0}$. This means that, if the photon's mean free path vanishes, the radiation gas is in local thermodynamic equilibrium with the fluid at all events, as one would expect.

\vspace{-0.2cm}
\subsubsection{Energy balance}
\vspace{-0.2cm}

Consider again equation \eqref{boltzmann}. Multiply both sides by $p^0$, integrate over all momenta, and use \eqref{granch}. The result is
\begin{equation}
\partial_\mu \int \dfrac{2d^3 p}{(2\pi)^3}p^\mu  \delta f_\mathbf{p} +\dfrac{c_R}{\tau} \int \dfrac{d^2 \Omega}{4\pi} \int_0^{+\infty} \delta T(t{-}\lambda,\mathbf{x}{-}\lambda\mathbf{\Omega}) \dfrac{e^{-\lambda/\tau}}{\tau} d\lambda=\dfrac{c_R \delta T}{\tau} \, .
\end{equation}
Combining this equation with the first line of \eqref{EoMsony}, we obtain
\begin{equation}
 \partial_t \delta T =-\dfrac{\lambda}{\tau} \bigg[\delta T - \int \dfrac{d^2 \Omega}{4\pi} \int_0^{+\infty} \delta T(t{-}\lambda,\mathbf{x}{-}\lambda\mathbf{\Omega}) \dfrac{e^{-\lambda/\tau}}{\tau} d\lambda  \bigg]  \, . \\
\end{equation}
The integral on the right-hand side can be expressed as a convolution of $\delta T$ with a retarded propagator. Specifically, if we define a displacement vector $\mathbf{r}=\lambda \mathbf{\Omega}$, and make the change of variables $\{\lambda,\mathbf{\Omega} \}\rightarrow \mathbf{r}$, the 3D volume form $d^2\Omega d\lambda$ becomes $d^3 r/|\mathbf{r}|^2$, and the integration domain $\mathcal{S}^2 \times [0,+\infty)$ becomes just $\mathbb{R}^3$. Therefore, we have
\begin{equation}\label{Ccinque}
 \partial_t \delta T(t,\mathbf{x}) =-\dfrac{\lambda}{\tau} \bigg[\delta T(t,\mathbf{x}) - \int_{\mathbb{R}^3}  \delta T(t{-}|\mathbf{r}|,\mathbf{x}{-}\mathbf{r}) \dfrac{e^{-|\mathbf{r}|/\tau}}{4\pi \tau |\mathbf{r}|^2} d^3 r  \bigg]  \, . \\
\end{equation}
This equation is identical to equation (7) of \citet{Spiegel1957}, except for one difference: the time retardation $t-|\mathbf{r}|$ in the propagation integral. It is the replacement $t-|\mathbf{r}|\rightarrow t$ that makes Spiegel's model unstable in relativity.

\vspace{-0.2cm}
\subsubsection{Effect of time retardation on dispersion relations}
\vspace{-0.2cm}

For completeness, let us also verify that the retarded time gives rise to the denominator $1{-}i\omega\tau$ in equation \eqref{fullrelat}. To this end, we only need to set $\delta T=e^{ikx^1-i\omega t}$ in \eqref{Ccinque}, which gives
\begin{equation}\label{Csei}
\omega =-i\dfrac{\lambda}{\tau} \bigg[1 - \int_{\mathbb{R}^3}  \dfrac{ e^{-ikr^1-(1-i\omega\tau) |\mathbf{r}|/\tau}}{4\pi \tau |\mathbf{r}|^2} d^3 r  \bigg]  \, . \\
\end{equation}
We see that, due to time retardation, there is a factor $1{-}i\omega\tau$ in the exponent. If we carry out the integral, we eventually recover equation \eqref{fullrelat}, with $1{-}i\omega\tau$ in the right place.

\newpage

\bibliography{Biblio}

\label{lastpage}

\end{document}